\documentclass[aps,prb,twocolumn, showpacs, superscriptaddress, showkeys]{revtex4}
\usepackage[colorlinks,bookmarksopen]{hyperref}
\usepackage{graphicx}
\usepackage{dsfont}
\usepackage{xspace}
\usepackage{amsmath}
\usepackage{epstopdf}

 \newcommand{\f}{\mathbf}                         
\newcommand{\txt}{\mathrm}                        % text in math mode
\newcommand{\ie}{\mbox{i.\,e.}\xspace}

\newcommand{\eg}{\mbox{e.\,g.}\xspace}
\newcommand{\eV}{\mbox{e\hspace{0.08mm}V}\xspace}          
        
\begin{document}

%\title{Comparison of calculated and experimentally determined band offsets}
\title{Determination of the valence band offset at cubic CdSe/ZnTe
type II heterojunctions: A combined experimental and theoretical approach}
%\author{JP Richters$^1$, D Mourad$^2$, L G\'erard$^1$, R Andr\'e$^1$, J Bleuse$^1$ and H.
%Mariette$^1$}

\author{Daniel Mourad}
\email{dmourad@itp.uni-bremen.de}
\affiliation{Institute for Theoretical Physics, University of Bremen,
Otto-Hahn-Allee 1, 28359 Bremen, Germany}

\author{Jan-Peter Richters}
\affiliation{CEA-CNRS group 'Nanophysique et Semiconducteurs', CEA-Grenoble, INAC, SP2M, 17 Rue des Martyrs, 38054 Grenoble CEDEX 9, France}

\author{Lionel G\'erard}
\affiliation{CEA-CNRS group 'Nanophysique et Semiconducteurs', Institut N\'eel-CNRS/Universit\'e Joseph Fourier, 25 Rue des Martyrs, 38042 Grenoble CEDEX 9, France}

\author{R\'egis Andr\'e}
\affiliation{CEA-CNRS group 'Nanophysique et Semiconducteurs', Institut N\'eel-CNRS/Universit\'e Joseph Fourier, 25 Rue des Martyrs, 38042 Grenoble CEDEX 9, France}

\author{Jo\"el Bleuse}
\affiliation{CEA-CNRS group 'Nanophysique et Semiconducteurs', CEA-Grenoble, INAC, SP2M, 17 Rue des Martyrs, 38054 Grenoble CEDEX 9, France}

\author{Henri Mariette}
\affiliation{CEA-CNRS group 'Nanophysique et Semiconducteurs', Institut N\'eel-CNRS/Universit\'e Joseph Fourier, 25 Rue des Martyrs, 38042 Grenoble CEDEX 9, France}

%\affiliation{CEA-CNRS group 'Nanophysique et Semiconducteurs'\\
%$^1$ CEA-Grenoble, INAC, SP2M, 17 Rue des Martyrs, 38054 Grenoble CEDEX 9, France\\
%$^2$ Institut N\'eel-CNRS/Universit\'e Joseph Fourier, 25 Rue des Martyrs, 38042 Grenoble CEDEX 9, France}


\begin{abstract}
We present a combined experimental and theoretical approach for the 
determination of the low-temperature valence band offset (VBO)
at CdSe/ZnTe heterojunctions 
with underlying zincblende crystal structure. 
On the experimental side, the optical transition of the type II interface
allows for a  precise measurement of the type II band gap. We
show how 
the excitation-power dependent shift of this photoluminescence (PL) signal 
can be used for any type II system for a precise determination of
the VBO.
On the theoretical side, we use a refined empirical tight-binding
parametrization in order
to accurately reproduce the band structure and density of states around the 
band gap region of cubic CdSe and ZnTe and then calculate the branch point
energy  (also known as charge neutrality level) for both materials. Because of
the cubic crystal structure and the small lattice mismatch
 across the interface, the VBO for the material 
system  under consideration can then be obtained from a charge neutrality
condition, in  good agreement with the PL measurements.
\end{abstract}

\maketitle

%%%%%%%%%%%%%%%%%%%%%%%%%%%%%%%%%%%%%%%%%%%%%%%%%%%%%%%%%%%%%%%%%%%%%%%%%%%%%%%

\section{Introduction}

 The knowledge of valence band offsets (VBOs) is
crucial for the design of
efficient semiconductor devices, especially when dealing with type II 
band gap alignment. In general, two possibilities exist for the band alignment 
of semiconductors. The most widely investigated are the so-called type I 
systems, in which the narrower gap material plays the role of the potential
well for
 both the 
electrons and holes, as its conduction band (CB) minimum/valence band (VB)
maximum lies at a lower/higher energy than the respective counterparts 
of the material with the larger gap. 
There exists, however, another group of semiconductor structures
known as type II structures, in which the band gap regions are staggered across
the 
 interface normal direction, \ie the  CB minimum and the  VB
maximum of one material are both at lower energies than the corresponding 
quantities
of the second material (the so-called type III or broken-gap alignments, where 
the CB minimum of one material dips even below the VB maximum of its
counterpart
can be considered a special case of  type II alignments).
%: The material 
%with the lower potential energy for electrons has the higher potential for 
%holes and vice versa. 
Thus for type II alignment, electron and hole wavefunctions are spatially 
separated across the interface. 
This separation gives rise to  relatively long carrier lifetimes and to a 
dependence of photoemission and photocurrent on the intensity of excitation, 
as well as on external electric and magnetic fields. Moreover, type II systems
have another important advantage in that they tend to suppress Auger 
recombination.\cite{zegrya_mechanism_1995} These properties, as well as other
that result from the type II  band alignment, provide unique opportunities for 
new optical properties, e.g. Aharonov-Bohm effects in type II quantum dots, 
\cite{kuskovsky_optical_2007, sellers_aharonov-bohm_2008}
and new potential applications, e.g. for solar cells.
\cite{zhang_quantum_2007,schrier_optical_2007}

There are many different experimental and 
theoretical methods to determine the VBO for various material combinations.
However, the spread of the calculated values is rather large, as are the
results of some experimental methods.  The calculation of the VBO 
 at the interface between two 
semiconductor systems has been an important topic for decades. Although 
first principle methods\cite{wei_calculated_1998} have become more feasible in
recent time, they often
still lack satisfactory quantitative agreement with experiments, especially 
for the electronic properties of semiconductors. Furthermore, they do not 
give insight in the underlying physics at those semiconductor junctions.
As a consequence, a lot of effort has been put into models that basically 
trace the band alignment back to the bulk properties of the constituents by
establishing a common reference level. While there exist many works of
different
levels of sophistication
which align the constituents at the junction to vacuum
levels,\cite{anderson_experiments_1962,van_vechten_quantum_1969,
van_de_walle_theoretical_1986,van_de_walle_band_1989}
another class of 
models which calculate the band offset from a charge neutrality condition seems
to be more promising
nowadays.~\cite{harrison_tight-binding_1986,
cardona_acoustic_1987,van_de_walle_universal_2003,schleife_branch-point_2009,
hoffling_band_2010,monch_branch-point_2011}

The VBO of zincblende CdSe/ZnTe heterojunctions is an important
parameter for the simulation of hetero- and nanostructures like colloidal type
II nanocrystals.~\cite{kim_type-ii_2003}
While there are several experimental methods known in order to determine 
the VBO,~\cite{shih_determination_1987, biswas_conduction_1990,
lang_measurement_1987} 
the most widely renowned work on CdSe/ZnTe was 
performed by Yu et al.\,who performed X-ray photoelectron spectroscopy (XPS) measurements and found the
VBO  to be $(0.64 \pm 0.07)$ \eV at room temperature.\cite{yu_measurement_1991}
The strongest limitation of 
XPS measurements is the often poor energy resolution, having an uncertainty 
of often several 100 meV. Another work by Gleim et
al.\,combines \textbf{k}-resolved valence- and core-level photoelectron
spectroscopy and 
gives a
similar result of $(0.6 \pm 0.1)$ \eV.~\cite{gleim_energy_2002} The direct
measurement of the optical
transition of a type II interface is another method which promises high 
precision and easy application. Mostly, this method has been used on multi
quantum wells made of the In-Al-As-Ga-Sb materials. 
\cite{dawson_staggered_1986, ostinelli_photoluminescence_2006,
glaser_determination_2006}

In this work, we present a combined theoretical and experimental 
approach to determine the VBO between CdSe and ZnTe and compare the 
calculations with the experimentally determined value. 
The theoretical part presents a refined approach for the calculation of 
the VBO. It is based on methods known from the literature, but has been
improved for a
better incorporation of the one-particle properties of the constituents by
using 
a customized empirical tight-binding parametrization.
The experimental part utilizes the excitation-dependent measurement of the 
photoluminescence (PL) signal of the spatially indirect type-II band-gap
between 
CdSe and ZnTe. We introduce a model to correctly extract the type-II band gap
from the obtained PL spectra in order to calculate the VBO.

%%%%%%%%%%%%%%%%%%%%%%%%%%%%%%%%%%%%%%%%%%%%%%%%%%%%%%%%%%%%%%%%%%%%%%%%%%%%%%%

\section{Experimental}

The samples were fabricated using a Riber 32P molecular beam epitaxy (MBE)
machine. 
First, a 200 nm thick layer of ZnTe is deposited on an InAs substrate under
 Zn-rich conditions, followed by 400 nm of CdSe under Se-rich conditions. 
The very low lattice mismatch between cubic CdSe and ZnTe (a$_\txt{CdSe}=
6.077$
\AA\, and a$_\txt{ZnTe}= 6.089$ \AA, $\Delta a \approx 0.2\%$ 
) \cite{park_molecular_2007}
allows the growth of metastable, cubic CdSe with very good crystal quality,
\cite{samarth_molecular_1990} as has 
been confirmed by X-ray diffraction (XRD) and PL measurements.

The PL measurements presented in this work were performed using a frequency 
doubled Coherent Mira Ti:sapphire fs-laser operating at 810 nm 
(doubled to 405 nm). The samples are placed in an Oxford liquid Helium cryostat
which is mounted on a Zeiss Axiovert 200 inverted microscope. The excitation
laser is
guided into the microscope using a multi-mode optical fiber. A Zeiss
20x microscope objective focuses the laser on the sample and collects the 
resulting photoluminescence (PL) signal. The spectral analysis of the PL 
is performed with a Jobin Yvon Triax 320 spectrometer using an 
thermo-electrically cooled Andor InGaAs CCD. 

%%%%%%%%%%%%%%%%%%%%%%%%%%%%%%%%%%%%%%%%%%%%%%%%%%%%%%%%%%%%%%%%%%%%%%%%%%%%%%%

\section{Theory}

%-------------------------------------------------------------------------
\subsection{Calculation of valence band offsets from the charge neutrality
            condition \label{subsec:calculationofvbo}}
%-------------------------------------------------------------------------

% The calculation of the valence band offset (VBO) $\Delta E_\txt{v}$ at the
% interface between two 
% semiconductor systems has been an important topic for decades. Although 
% first principle methods have become more feasible in recent time, they still
% lack satisfactory quantitative agreement with experiments, especially 
% for the electronic properties of semiconductors. Furthermore, they do not 
% give insight in the underlying physics at those semiconductor junctions.
% As a consequence, a lot of effort has been put into models that basically 
% trace the band alignment back to the bulk properties of the constituents by
% establishing a common reference level.
% 
% While there exist many works of different levels of sophistication
% which align the constituents at the junction to vacuum levels \cite{bla},
% another class of 
% models which calculate the band offset from a charge neutrality condition
%seems
% to be more promising.
When two semiconductors A and B are brought together to form a common
interface, the electronic structure will be altered, as the translational
invariance is broken in the direction of the interface normal. 
Localized interface-induced states with an energy
either in the A or B gap or in the coinciding gap of the
constituent semiconductors can be
shown to carry net charge density across the junction, thus inducing an 
interface dipole.~\cite{flores_energy_1979}
These states can be interpreted as Bloch-like bulk states of one
material that decay exponentially into the other
material when crossing the 
interface.~\cite{monch_electronic_2004}

Tersoff argued in
Ref.\,\onlinecite{tersoff_theory_1984} that a filled interface state results in
local excess 
charge density, proportional to its conduction character, while an unoccupied 
state leads to a local charge density deficit proportional to its valence
character. At a certain
energy, the 
spectral character of an interface state is of equal CB and
VB origin, \ie  neither donor- nor acceptorlike. This energy
can thus be identified as the charge neutrality level of the interface
states and is commonly called the branch point (BP) energy $E_\txt{BP}$.
Therefore, the orientation of the overall resulting dipole depends
on the relative energetic position of the Fermi level to $E_\txt{BP}$. As the
interface states can be traced back to A or B bulk states that reach across
the common interface, the BP energy can be seen as an interface-orientation
 dependent intrinsic property for the
given  A and B material, respectively.

Similar to other approaches which align materials to the charge neutrality
levels of hydrogen impurities (see \eg
Ref.\,\onlinecite{van_de_walle_universal_2003}), the BP energy can
 serve as a common reference level for band alignment:
The only case where no resulting dipole 
is left would be the lineup where the BPs of the A and
B material coincide (the so-called \textit{canonical lineup}), as all
interface dipoles will then cancel out.
On the other hand, polarization effects will screen charge transfer effects
with a characteristic
dielectric constant $\varepsilon$, \ie counteract any deviation from this
canonical lineup.
As $\varepsilon$ is relatively high for most compound semiconductors 
($\varepsilon \approx 10\ldots20$), the overall charge transfer is effectively
reduced and the canonical lineup condition 
will approximately be fulfilled for most isovalent material combinations,
regardless of the stochiometry-related polarity of the specific interface
(see \eg Ref.\,\onlinecite{lambrecht_interface-bond-polarity_1990} for a
detailled
analysis). If we measure the BP energy $ E_\txt{BP}^\txt{A/B}$ of each
constituent from its VB edge, respectively, the VBO
can then be calculated as
\begin{equation}
\Delta E_\txt{v} \approx E_\txt{BP}^\txt{B} - E_\txt{BP}^\txt{A}
\end{equation}
\label{eq:lineupcondition}
with an accuracy of $\sim 0.05$ \eV.
\cite{tersoff_theory_1984,
monch_empirical_1996}
Therefore, the problem of band alignment has approximately been reduced to the
separate calculation of $E_\txt{BP}$ for each material. 

%-------------------------------------------------------------------------
\subsection{Calculation of branch point energies}

%-------------------------------------------------------------------------

According to Allen,
\cite{allen_greens_1979} the cell-averaged real-space
Green's function for the propagation across a surface or interface with normal 
vector parallel to the direct lattice vector $\f{R}$ is given as
\begin{equation}
G^\f{R}(E) 
%   =  \int d^3r \sum \limits_{n \f{k}}
%     \frac{\psi^*_{n \f{k}}(\f{r}) \psi_{n
%\f{k}}(\f{r}+\f{R})}{E-E_{n}(\f{k})}
    =  \sum\limits_{n \f{k}}
    \frac{e^{i \f{k}\cdot\f{R}}}{E-E_{n}(\f{k})},
\label{eq:greensfunction}
\end{equation}
where $E_n(\f{k})$ is the band structure, $\f{k}$ the wave
vector in the first Brillouin
zone (BZ) and $n$ the band index.
In accordance with its definition in the preceding subsection, the BP is then
given as the energy where the bulk CBs and
VBs contribute in equal parts to $G^\f{R}(E)$. Although this method 
allows for the calculation of interface-orientation dependent BP energies, it
can be rather tedious numerically, as one has to converge $G^\f{R}(E)$ for 
multiples of the smallest $\f{R}$ for each interface orientation to project
out the relevant contributions.\cite{tersoff_schottky_1984}

An alternative is the usage of interface-averaged approximations for the
calculation of $E_\txt{BP}$. Similar to Cardona and Christensen,~\cite{
cardona_acoustic_1987}
Schleife et
al. generalized previous approaches
\cite{flores_energy_1979, tersoff_schottky_1985} and
calculated the BP as a BZ average of the midgap
energy:~\cite{schleife_branch-point_2009}
\begin{equation}
 E_{\txt{BP}} \approx \frac{1}{2N_\f{k}} \sum\limits_{\f{k}} \left[
\frac{1}{N_\txt{CB}}  \sum\limits_i^{N_\txt{CB}} E_\txt{CB}^{i}(\f{k}) + 
\frac{1}{N_\txt{VB}} \sum\limits_j^{N_\txt{VB}} E_\txt{VB}^j(\f{k})
\right].
\label{eq:bzaverage}
\end{equation}
Here, $N_\f{k}$ is number of $\f{k}$-points and $N_\txt{CB}$, $N_\txt{VB}$ are
the numbers of included CBs and VBs with dispersions
$E_\txt{CB}^{i}(\f{k})$ and $E_\txt{VB}^{j}(\f{k})$, respectively.
The $\f{k}$-sample
can be confined to the irreducible wedge of the corresponding BZ, as no
interface orientation breaks the symmetry properties of the bulk dispersion,
at the expense of losing the orientational dependence.
When a sufficiently
dense $\f{k}$-sample is available, Eq.\,(\ref{eq:bzaverage}) can be reduced to
an integration over sufficiently smooth densities of states (DOS),
\begin{equation}
\label{eq:dosbzaverage}
 E_{\txt{BP}} \approx
 \frac{1}{2} \int dE\,E \left[
    \frac{1}{N_\txt{CB}}  \sum\limits_i^{N_\txt{CB}}
  g_\txt{CB}^{i}(E)
  + \frac{1}{N_\txt{VB}} 
   \sum\limits_j^{N_\txt{VB}} g_\txt{VB}^j(E) \right],
\end{equation}
where the band-resolved DOS (normed to unity) is given as
\begin{equation}
g^i(E) = \frac{1}{N_\f{k}} \sum_\f{k} \delta \left[ E-E^{i}(\f{k})\right].
\end{equation}

In contrast to the Green's function method, Eq.\,(\ref{eq:dosbzaverage}) is
also applicable to cases where the BP lies outside the
band gap, as for example in InAs or InN. The pinning of the Fermi
level near the branch point then results in electron accumulation at free
surfaces, which can be observed by means of high-resolution
electron-energy-loss spectroscopy (HREELS).~\cite{noguchi_intrinsic_1991,
mahboob_origin_2004, piper_electron_2006}

To sum it up, the calculation of BPs 
requires the choice of an appropriate input band
structure [when using Eq.\,(\ref{eq:greensfunction})] or DOS
[Eq.\,(\ref{eq:dosbzaverage})] and in practice
the choice of a proper subset of contributing bands around the band gap region.
For realistic band structures, the occurrence of Van Hove singularities due to
the vanishing slope of non-(quasi-)crossing bands at the BZ boundaries will
further complicate the problem, as dense $\f{k}$-point samples are required to
represent such
kinks in the DOS. On the other hand, the typically shallow
dispersion at the zone faces
is of crucial influence to the branch point position, as localized levels
sample large $\f{k}$-space regions. The usage of simple effective mass or
$\f{k}\cdot\f{p}$ models is suitable for describing direct optical
transitions near the $\Gamma$-point, but of little help here due to the
erroneous dispersion for large wave vectors. Consequently, also the common 
use of nonlinearly-spaced $\Gamma$-centered  meshes in more costful (\eg
quasiparticle) calculation schemes could in some cases counteract the accuracy
of the BP
calculation, as they typically give the largest $\f{k}$-density in the least
contributing BZ region. Furthermore, it is well known that even
highly-sophisticated ab initio calculation schemes still struggle with the
quantitative reproduction of electronic features of
semiconductors.~\cite{rinke_combining_2005}

As trade-off, empirical
tight-binding models (ETBMs)\cite{slater_simplified_1954} seem to
be well-suited,~\cite{monch_empirical_1996} as they allow for a realistic
dispersion throughout the whole BZ, combined with a flexible parametrization
scheme. Unfortunately, most ETBM parametrizations are also optimized  to
reproduce the zone center properties rather than those at the BZ boundaries.

To overcome these adversities, there exists a parametrization scheme by
Loehr
\cite{loehr_improved_1994} that fits the band
structure of zincblende type crystals to the $X$-point energies within a
small basis set of Wannier-like
bonding orbitals on the Bravais lattice sites of the underlying crystal.
For
the wurtzite structure, a similar approach even allows
for  the fit of one CB and three VBs to almost all high-symmetry points
in the hexagonal BZ, as the low crystal symmetry
allows for a large number of independent
parameters.~\cite{mourad_multiband_2010}

Although the ETBM by Loehr is well suited for the calculation of electronic
and optical properties for a wide range of material systems and
geometries,~\cite{mourad_band_2010,marquardt_comparison_2008,
schulz_multiband_2009,mourad_multiband_2010,
mourad_multiband_2010-1,mourad_theory_2012,schulz_tightbinding_2011} we found
it to be of lower accuracy in the specific case of BP
calculations for zb-CdSe and especially ZnTe. Due to the large spin-orbit
coupling constant (the
spin-orbit splitting of ZnTe at $\f{k}=\f{0}$ is about 1
{\eV}),~\cite{adachi_handbook_2004} the $L$-point energies 
do not coincide well with literature values\cite{adachi_handbook_2004} when not
fitted explicitly. In this paper, we
will therefore
follow a modified approach within the same basis set. We fit one CB and
three VBs, namely the heavy hole (HH), light hole (LH) and split-off band,
of the zincblende band structure  to the zone center masses and the
energies at
$\Gamma$ and $X$, and additionally fix the position of the CB, HH and LH VB at
$L$. More details are given in Appendix
\ref{sec:tightbindingparametrization}.

%-------------------------------------------------------------------------
\subsection{Experimental determination of valence band offsets from type-II
            PL spectra}
%-------------------------------------------------------------------------

In order to obtain the VBO from  photoluminescence measurements
of the CdSe/ZnTe type-II interface, 
one has to take into account the special behavior of the charge carriers at 
the interface. Due to the type II band alignment, photo-generated electrons 
are confined in CdSe, while the holes are confined in ZnTe. Their attractive
Coulomb force leads to an accumulation at the interface which causes an 
electric field $E_\txt{field}$  that is proportional to the population of
carriers
 $n$ at the type-II interface, similar to a plate capacitor. This field can be
described by 
a band-bending, which can be approximated for a planar type II interface as 
a triangular potential well.~\cite{lui_exact_1986}

The recombination rate $\Gamma$ (a mix-up with the notation for the BZ center
can be excluded in the context of this section) of charge carriers
depends 
of course on the wave-function overlap between electrons and holes at the 
interface. At a type II interface, this overlap becomes a function of the
charge carrier density at the interface. 
In Ref.\,\onlinecite{shuvayev_dynamics_2009} , Shuvayef et al.\,present
the charge carrier dynamics of a type II system obtained by a numerical 
solution of the self-consistent Schr\"odinger-Poisson system of equations. 
Their numerical results fit the experimental results very nicely. 
In this work, we propose a model that displays a comparably good agreement
between theory and experiment, but by using an analytical solution of the 
carrier dynamics of a type II interface.

To obtain the dependency of the  recombination rate $\Gamma$ on the carrier 
population, we numerically solved the Schr\"odinger equations to get the
overlap between electron and hole wavefunctions for different electric fields.
The overlap has been found  to be proportional to the 
electric field $E_\txt{field}$ at the interface,
\cite{andre_unpublished_2012} 
therefore the recombination rate $\Gamma$ becomes proportional to the carrier
density $n$. Then we can introduce a constant 
$\gamma$ such that $ \Gamma = \Gamma(n) = \gamma \, n$.

Next to be considered is the type of emission observed at a type II interface.
The usual case for (in $\f{k}$-space) direct  or indirect band gap
semiconductors at low temperatures 
and low excitation conditions is the observation of excitonic emission. With
 strongly 
increasing pumping power, the excitonic emission will evolve into a 
band-to-band emission (electron-hole plasma). For an emission from a type II
interface this  behavior is different, because the formation of excitons 
is hindered at low excitation  conditions: Degani and Farias calculated in 
Ref.\,\onlinecite{degani_exciton_1990}  the binding energy of an exciton at
the 
type II interface of AlAs/GaAs under the assumption of infinite potential 
barriers for electrons and holes at the interface. They found the binding 
energy to increase with excitation power, but to be zero for excitation 
densities below a certain threshold. For finite barriers they estimate a 
general increase of the binding energy, leading to a lower threshold
excitation 
density for the formation of excitons. Therefore, in our results, one has to 
expect a band-to-band transition (with a bimolecular recombination behavior) 
to be the main origin of the type II PL with
an excitonic feature eventually coming up with increasing pumping power.

Starting from a rate equation for band-to-band recombination one can derive 
a description of the type II recombination dynamics by including the above
mentioned
carrier dependent recombination rate $\gamma \, n$ under the assumption of an
equal electron and hole population ($n = p$):
\begin{equation}
\frac{dn}{dt} = \alpha \, P - (\gamma \, n) \, (n \, p) = \alpha P -
\gamma \, n^3 \ .
\label{eqn:rateeqn}
\end{equation}
Here, $P$ is the optical pumping power and $\alpha$ a proportionality factor. 
The solution to this equation is a hyperbolic function. Its time derivative
itself gives the
observable PL-intensity $I = {dn}/{dt}$ which very nicely 
describes the experimentally found non-exponential decay behavior. This is 
the first of two typical features for a type II emission,
\begin{equation}
I(t) = \frac{\gamma \, n_0^2}{(1+\gamma \, n_0 \, t)^{3/2}},
\label{eqn:decay}
\end{equation}
with $n_0=n(t=0)$.
Solving the steady-state case of Eq.\,(\ref{eqn:rateeqn})  for $n$ gives
the excitation power dependence of the number of generated charge carriers:
\begin{equation}
n = \left( \frac{\alpha}{\gamma} P \right)^{1/3}.
\end{equation}
As described above, their attractive Coulomb interaction leads to an 
accumulation of charge carriers along the interface, creating a band bending
which 
can be described as a triangular potential well.~\cite{lui_exact_1986} 
In such a potential the 
charge carriers experience a confinement $E_c$ which is a function of 
the electric field which itself is proportional to the number of charge 
carriers $E_\txt{c} \propto E_\txt{field}^{2/3} \propto n^{2/3}$.
Accordingly, the
dependence of the confinement energy on the optical pumping power is given by 
$E_c \propto P^{2/9}$.
This confinement leads to a blueshift of the PL signal which is the second 
typical feature of a planar type II interface. The position of the PL signal
$E(P)$ 
can then be fully described by
\begin{equation}
E(P) = E_0 + \beta \, P^{2/9},
\label{eqn:shift}
\end{equation}
where $E_0$ is the zero-excitation, spatially indirect band gap of the type II 
junction and $\beta$ a proportionality factor. The decay dynamics of the 
PL emission at a type II interface strongly resembles the behavior 
of an electron-hole plasma, but can be observed well below the Mott transition 
and show a much longer lifetime.
 
%%%%%%%%%%%%%%%%%%%%%%%%%%%%%%%%%%%%%%%%%%%%%%%%%%%%%%%%%%%%%%%%%%%%%%%%%%%%%%

\section{Results \label{sec:Results}}

%-------------------------------------------------------------------------
\subsection{ETBM band structure of zb-CdSe and ZnTe}

%-------------------------------------------------------------------------
\begin{figure*}
\begin{center}
\includegraphics[width=\linewidth]{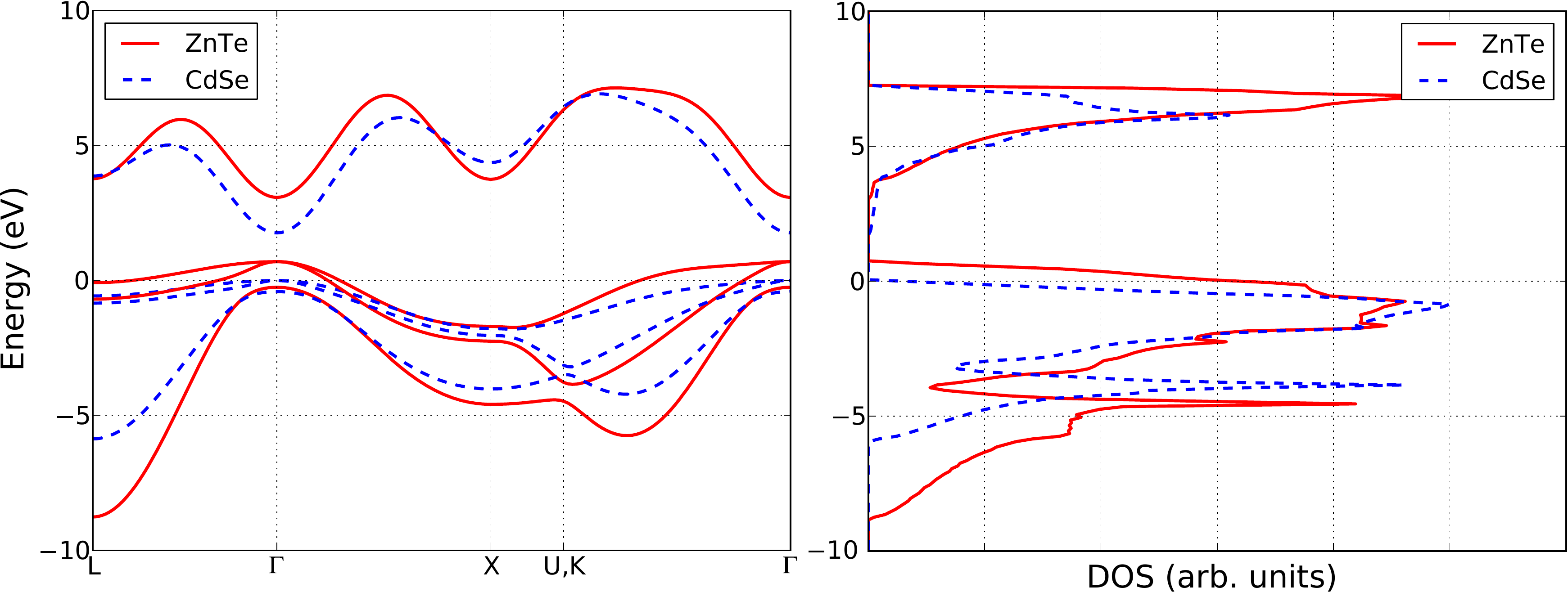}
\end{center}
\caption{(Color online) Band structures (left) and
DOS (right) of zb-CdSe
and ZnTe as obtained in our ETBM. Note that we already
use the later calculated VBO of $E_\txt{v} \approx 0.7$\,{\eV}  in these
plots.}
\label{fig:bs_and_dos}
\end{figure*}

Figure \ref{fig:bs_and_dos} shows the band structure and the DOS of
zb-CdSe and ZnTe as obtained in our ETBM. Due to time inversal
symmetry, all bands carry an additional twofold degeneracy throughout the
whole BZ. More details on the parametrization and the input parameters can be
found in the appendix of this work.

The fundamental influence of the spin-orbit
coupling can easily be identified by the large energetic separation of the
split-off band from the degenerate HH/LH at $\Gamma$.
Also, the energetic regions with "flat`` bands that contribute significantly
to the DOS integrals and therefore also to the
position of the BP can be located, e.g. around the X point for the VBs. 

As our ETBM as given in the appendix does not fix the split-off VB at $L$, it
dips relatively far down there for ZnTe. Although it is also possible to fit
the ETBM
to this value, we are then left with a worse reproduction of
the
dispersion of this band at $\Gamma$ (not shown here). Due to included
band-mixing effects, this also implicitly influences the HH/LH VB dispersion
and can lead to erroneous curvatures along $\Gamma$--$X$ for ZnTe. We found
the present parametrization to be most suitable
for a good agreement of the DOS in comparison to literature data
(see \eg Ref.\,\onlinecite{adachi_handbook_2004}), as the overall shape is then
very
well reproduced besides a small "tail`` of the lowest ZnTe VB on the low energy
side. Its influence
has been included in the error range for the calculated BPs and VBOs 
by comparing the respective results from either parametrizations.

In case of zb-CdSe, this problem does not occur. Here, the
dispersion of the split-off band is well reproduced throughout the BZ
including the zone boundaries
without an additional fit to $L$ when compared to available literature
data.~\cite{adachi_handbook_2004}

\begin{figure}
\begin{center}
\includegraphics[width=\linewidth]{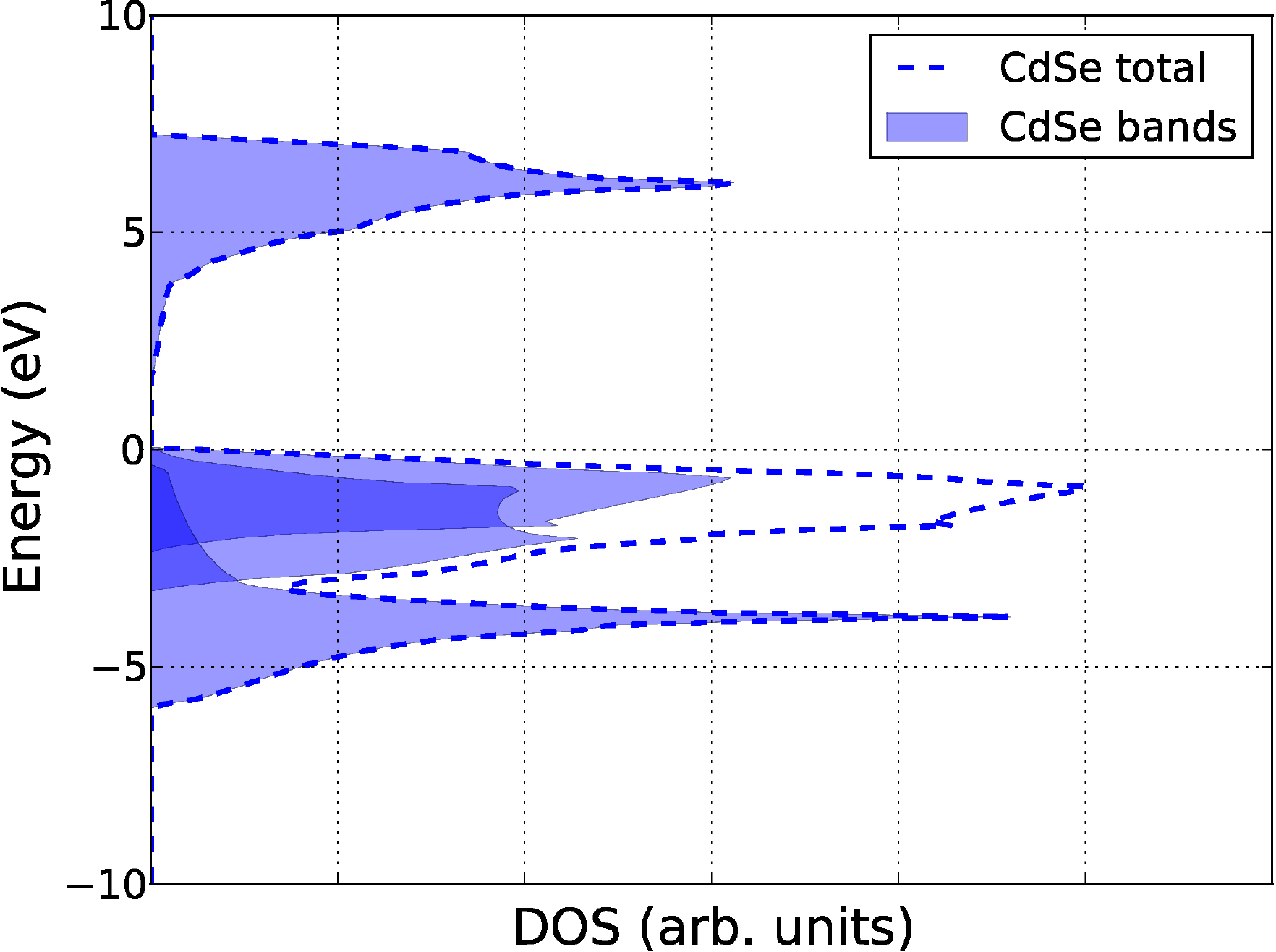}
\end{center}
\caption{(Color online). Band-resolved DOS of zb-CdSe.}
\label{fig:cdse_banddos}
\end{figure}

%-------------------------------------------------------------------------
\subsection{Calculation of branch points and valence band offsets of
zb-CdSe/ZnTe heterojunctions}
%-------------------------------------------------------------------------

As Schleife et al. pointed out in
Ref.\,\onlinecite{schleife_branch-point_2009}, the number of used bands can
introduce a relevant uncertainty when calculating the BP energies. 
Figure \ref{fig:cdse_banddos} exemplary shows the band-resolved DOS of
zb-CdSe. As the bandwidths of all bands are of comparable magnitude, we will
use one CB and all three VBs per spin direction in our BP calculations. This
also agrees well with the notion that in the zincblende structure the bands
around the gap are mainly formed from one $s$ orbital of the cation and three
$p$ orbitals of the anions on each unit 
cell.~\cite{schulz_tight-binding_2005}\footnote{In
similar III-V and III-nitride compound semiconductors with zincblende
structure, the split-off band shows a much larger dispersion with large
contributions far below the VB edge, so that it should likely be dismissed when
using only one CB per spin direction (see
Ref.\,\onlinecite{schleife_branch-point_2009}).}

The branch point calculation with the Green's function approach,
Eq.\,(\ref{eq:greensfunction}), turns out not to be suitable for the material
system under consideration, as  satisfactory convergence of
$G^{\f{R}}(E)$ could not be achieved for either material, regardless of the
interface orientation. When using the approximate formula
(\ref{eq:dosbzaverage}), it turns out that the BP of zb-CdSe lies slightly
above
the CB edge, so that Eq.\,(\ref{eq:greensfunction}) is not applicable. In case
of
ZnTe, it is most likely a numerical problem, as the multiple kinks in the DOS
of the VBs  render $G^{\f{R}}(E)$ extremely sensitive to the
$\f{k}$-resolution. Similar problems are reported in the
literature.~\cite{schleife_branch-point_2009} In contrast, convergence for the
branch point energy $E_\txt{BP}$ is easily reached with
Eq.\,(\ref{eq:dosbzaverage}). 

\begin{table}
\centering
\caption{Resulting branch point energies $E_\txt{BP}$ from our
specific ETBM with the BZ average approach. The literature values  were taken
from
Ref.\,\onlinecite{monch_empirical_1996}. All values are in \eV. The two
decimals are given for the sake of comparison and do not reflect the overall
accuracy of the calculations  (see text for
details).}
\label{tab:bpenergies}
\begin{tabular}{c|c|c}
\hline
\hline
 Material & $E_\txt{BP}$ (This work) & $E_\txt{BP}$
(Literature)\\
\hline
zb-CdSe & 1.83 & 1.53\\
zb-ZnTe & 1.09 & 0.73, 0.84, 1.00\\
\hline\hline
\end{tabular}
\end{table}

The results can be found in Table \ref{tab:bpenergies}, together with
literature data  from a comprehensive work from
M\"{o}nch.~\cite{monch_empirical_1996}
As mentioned above, our  BP energy  $E_\txt{BP} = 1.83\,\txt{\eV}$ for zb-CdSe
lies
above the low-temperature CB edge at $1.76\,\txt{\eV}$, while the available
literature value lies slightly below the CB minimum. 

Nevertheless, taking into account the multiple sources of ambiguities in the BP
calculations
[\eg uncertainties in the input parameters, the band structure
parametrization and the loss of the slight, but present directional dependence
of $E_\txt{BP}$ in
Eq.\,(\ref{eq:dosbzaverage})], we estimate the accuracy of the values to
$\pm0.1\,
\txt{\eV}$, so that we cannot decisively conclude whether the
BP really lies in the CB until suitable experimental evidence will be given,
e.\,g.\,on electron accumulation on CdSe surfaces.

%As the zincblende modification of CdSe is
%metastable, literature  values should refer to the
% wurtzite phase when no further information is given.
The result
from M\"{o}nch also refers the cubic modification, but it has been calculated
with
a different tight-binding parametrization based on multiple previous
works.~\cite{fischer_average-energy--configuration_1972,
harrison_new_1981,vogl_semi-empirical_1983} Specifically, the there employed
tight-binding matrix elements need to be rescaled with a heuristic factor in
order to satisfactorily reproduce the electronic structure of chalcogenides.
In addition, it is
also not obvious how a change in temperature affects these values, as the band
structure will in general be temperature-dependent. On the other hand, the
lack of reliable material data for metastable materials and their
temperature dependence adds another source of
uncertainty also on our side.

Similarly, the available literature values for ZnTe are all smaller than our
value $E_\txt{BP} = 1.09$ \eV. The first and the second value, however,
originally stem from LMTO and LAPW calculations, respectively, and will
therefore most likely suffer from the well-known issues with quantitative
predictions in these models.

Altogether, the charge neutrality condition leaves us with an estimated VBO of 
\begin{equation}
\Delta E_\txt{v}\approx E_\txt{BP}^\txt{CdSe} -
E_\txt{BP}^\txt{ZnTe} = (0.7 \pm 0.2)\,\txt{\eV}. 
\end{equation}
The quantitative influence of corrections due to non-vanishing interface
dipoles~\cite{monch_electronic_2004} can be estimated from the
electronegativity values from Miedema et al.~\cite{miedema_cohesion_1980} and
turns out to be small enough to be neglected (see Appendix
\ref{sec:dipolecontribution} for details). Further corrections could be
expected from the absence of a common atom across the interface, as the
inevitable presence of Cd--Te and Zn--Se bonds in one layer could in principle
introduce an additional confinement
potential.~\cite{priester_role_1994,gurevich_observation_2011}
This effect would also introduce an additional directional dependence, as has
been
reported in Ref. \onlinecite{su_polarized_2002}.

Nevertheless, the detailed theoretical studies of Lambrecht and
Segall\cite{lambrecht_interface-bond-polarity_1990} at the
example of polar InAs/GaSb interfaces come to the conclusion that for
isovalent interfaces without common anion no mentionable electric
field will result from these additional bonding configurations.
This is also in accordance with the studies of Priester et al.\,, who
explicitly
quantified the effect of an additional interface layer on the band
offset of the Al$_{x}$In$_{1-x}$As/InP heterojunction by means of
self-consistent tight-binding calculations.~\cite{priester_role_1994}

As a generalization, Lambrecht and Segall
find that details of the interface bonding configuration are only
relevant for nonisovalent heterojunctions (like II-VI/IV). Then the polarity
of the interface can play an
important role and additional dipole corrections can be relevant.
However, these consideration were made for cubic materials. In
systems with lower crystal symmetry, the direction-dependent bonding geometry
and stoichiometry of different surfaces might also play a role in case of
isovalent heterojunctions.

%-------------------------------------------------------------------------
\subsection{Experimentally determined valence band offset}
%-------------------------------------------------------------------------

\begin{figure}[htb]
\begin{center}
    \includegraphics[width =  \linewidth]{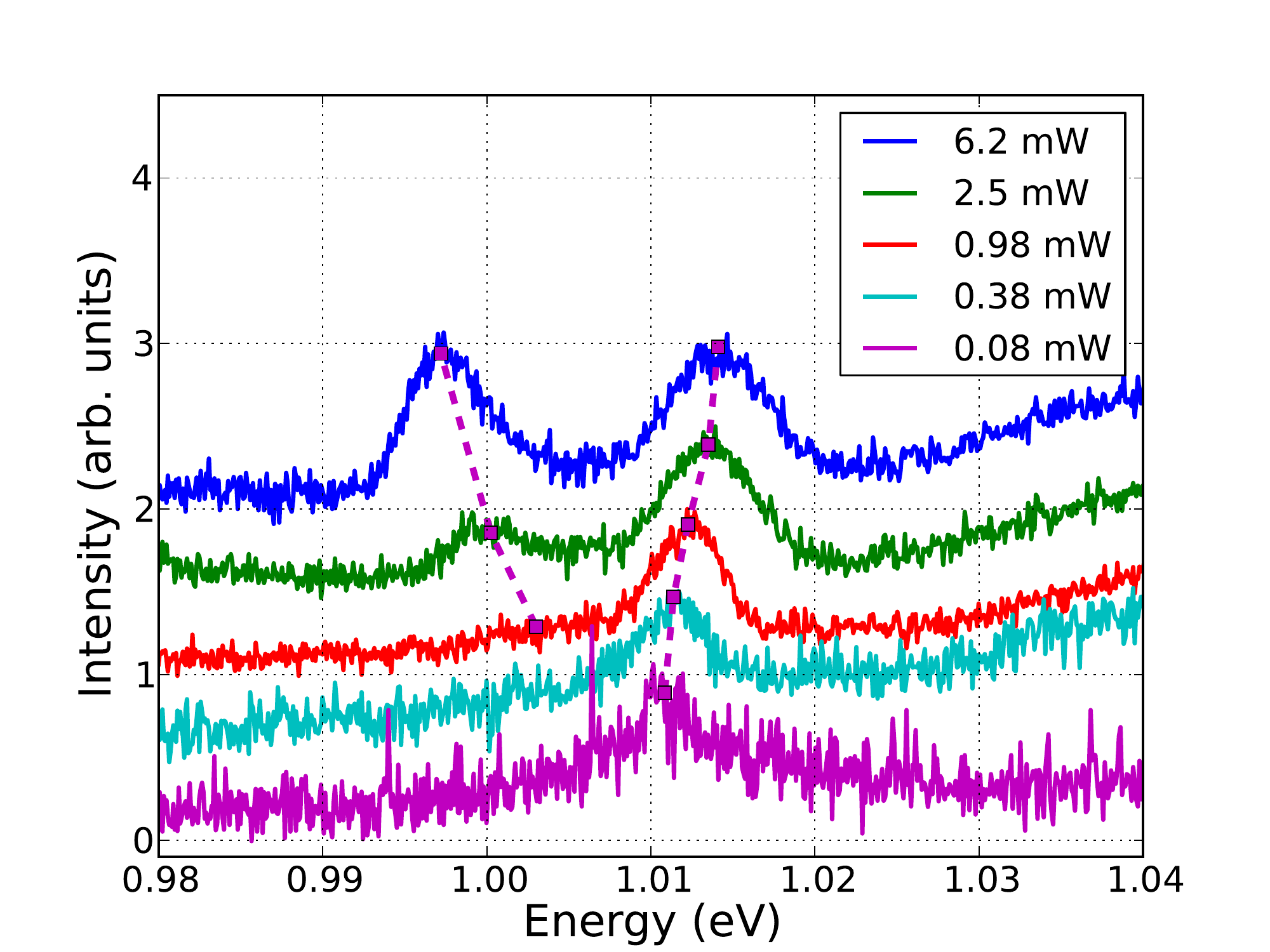}
\caption{(Color online). PL spectra of the type II interface obtained under
various excitation powers 
at 6 K. The higher energy
peak  can be attributed to the type II interface recombination (showing a 
blueshift with increasing excitation power). The lower energy peak is
probably 
the free exciton at the type II interface (showing a net redshift due to the
influence of the exciton binding energy).}\label{fig:spectra}
\end{center}
\end{figure}
Figure \ref{fig:spectra} displays the PL signal from the type II interface
between CdSe and ZnTe. This spatially indirect transition is 3--4 orders of
magnitude weaker in comparison to the direct transitions in CdSe and ZnTe. 
The spectrum consists of two peaks, one which is visible at all excitation
densities, and one which only becomes visible for large excitation densities
and is 
positioned on the lower energy side of the first peak.

\begin{figure}[htb]
\begin{center}
    \includegraphics[width =  \linewidth]{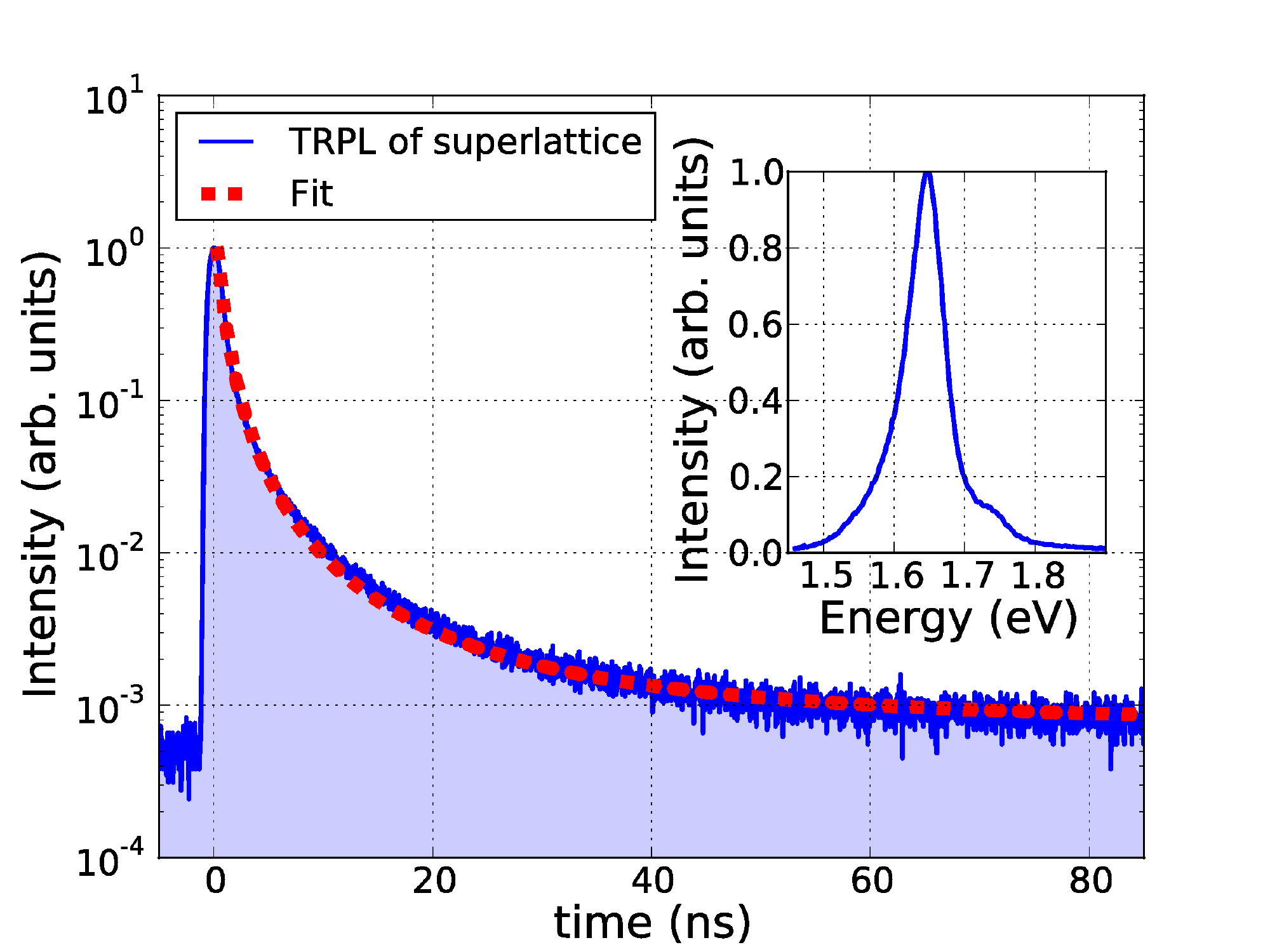}
\caption{(Color online). Decay behavior of the type II PL signal from a
superlattice fabricated
of 80 periods of 15 monolayers CdSe / 7 monolayers ZnTe. The dotted red curve
is a fit
according to Eq.\,(\ref{eqn:decay}). The signal has a very slow decay compared
to the direct transitions in CdSe and ZnTe, respectively ($\tau <$ 2 ns, not
shown).
The inset displays the time-integrated spectrum of this superlattice. The
strongest 
peak at 1.65 eV is the PL of the type II
transition.}\label{fig:PL-superlattice}
\end{center}
\end{figure}

The higher energy peak is  the type II band-to-band transition, which can 
be identified by its  excitation dependence. The PL of a planar type II
interface shows a blueshift [as described in Eq.\,(\ref{eqn:shift})] 
and a hyperbolic decay behavior. The blueshift is clearly visible in Fig.\,
\ref{fig:spectra}, but the PL signal of the samples presented in this work was
too weak to be investigated with time-resolved methods. However, we were able
to create  superlattices of CdSe/ZnTe in which the type II transition is
shifted into the visible region of the spectrum. Such samples show a blueshift
of the PL signal and a hyperbolic decay of the PL signal according to Eqns.\,
(\ref{eqn:shift}) and (\ref{eqn:decay}), respectively, as is displayed in
Fig.\,\ref{fig:PL-superlattice} . 

The  lower energy peak is only visible above a certain excitation 
density threshold. There are several hints that this line is the PL of the
free exciton at the type II interface: The decay dynamics of the free exciton 
are slightly different from the band-to-band transition and result in a 
squared-hyperbolic (bimolecular-like) decay and a shift of the PL position 
$E(P) = E_0 + \beta \, P^{1/3} + E_\txt{bind}(P)$ which includes the binding 
energy of the exciton at the type II interface $E_\txt{bind}$. The binding
energy 
is itself dependent on the pumping power and can be described by a 
power-law $E_\txt{bind} \propto  P^{0.12}$ (extracted from Fig.\,2 
in Ref.\,\onlinecite{degani_exciton_1990}).  This results in a net redshift of this
line with 
increasing excitation density.

When comparing the
calculated binding energy with 
the energy splitting between  the two peaks found in our measurements, we can
report 
 a similar power-law behavior $ E_\txt{bind}(\txt{CdSe/ZnTe}) \propto P^{0.08
\pm 0.1}$ with a
maximum binding energy of $E_\txt{bind} = 17$ meV at the highest excitation
power. Of 
course, there are very few data points in our measurements, leaving room for 
improvements. Further indication for the exciton being the origin of this line
lies
 in the excitation density dependent line intensity. 
The intensity of the band-to-band transition line (showing a blueshift) 
increases sub-linearly with the excitation density ($\propto P^{0.7 \pm
0.1}$), 
while the exciton line increases linearly with the excitation power ($\propto
P^{1.0 \pm 0.1}$).
This is typical for an excitonic emission. The sub-linear increase of the 
band-to-band transition could be explained by a strong nonradiative component 
for this transition, for example interface defects or Auger recombination. 
The summed intensity of both lines is still sub-linear. The last and strongest
hint
towards the exciton being the origin of the lower energy line is its eventual 
appearance above a certain threshold as it was predicted by Degani and Farias.
~\cite{degani_exciton_1990}

\begin{figure}[htb]
\begin{center}
\includegraphics[width = \linewidth]{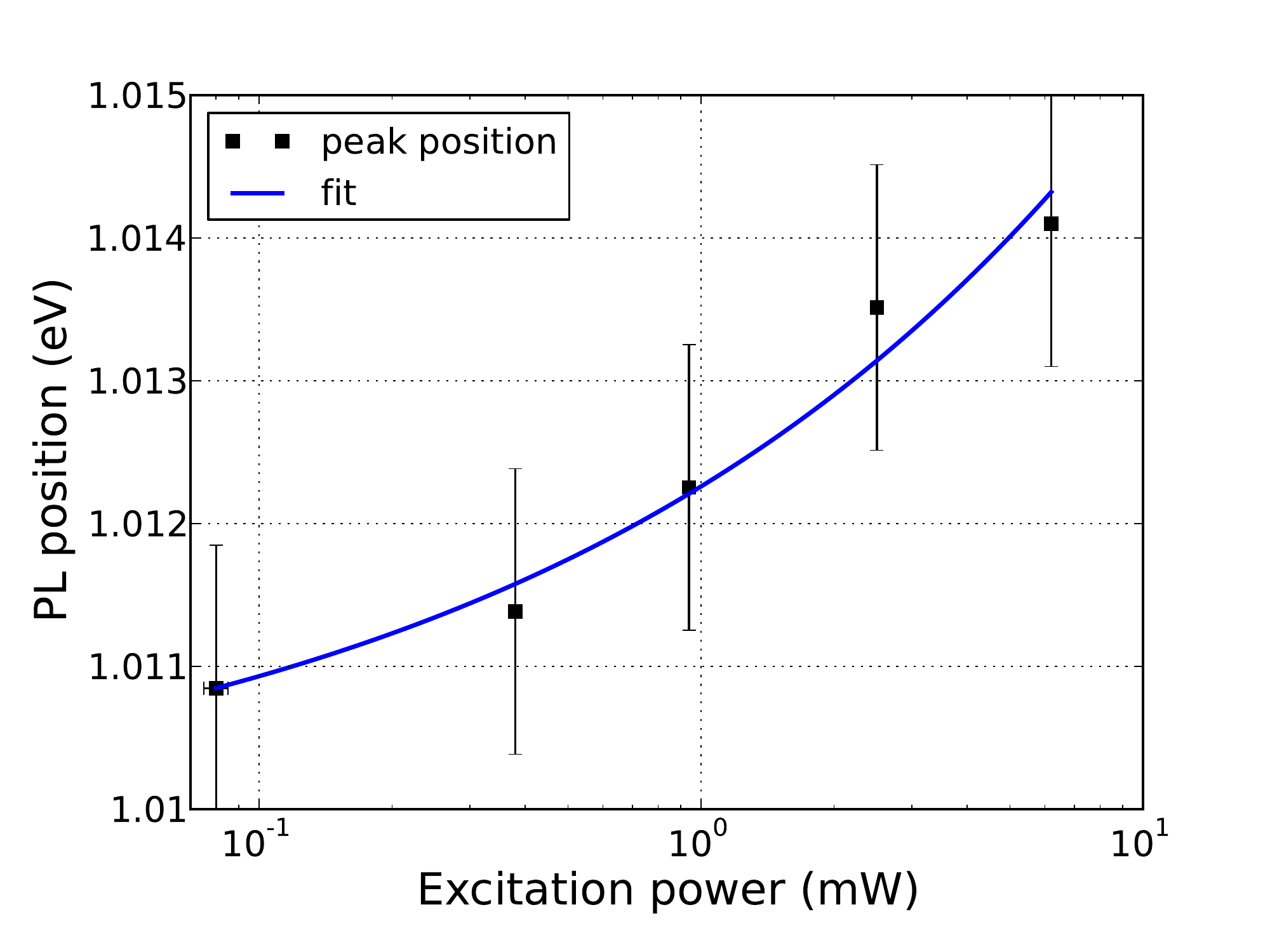}
\caption{(Color online). Shift of the type II band-to-band emission with
increasing excitation
power. The measurement data is shown as black squares. The curve is a power-law
fit $E = E_0 + P^{2/9}$ according to Eq.\,(\ref{eqn:shift}). The determined 
zero-excitation band-gap is $E_0 = (1.01 \pm 0.01)$ \eV.}\label{fig:shift}
\end{center}
\end{figure}

Figure \ref{fig:shift} shows the extracted blueshift of the band-to-band
emission 
with the excitation power. The curve in Fig.\,\ref{fig:shift} is a
fit according to Eq.\,(\ref{eqn:shift}). The zero-excitation type II band-gap
$E_0$ therefore is found to be 
\begin{equation}
E_0 = (1.01 \pm 0.01)\ \txt{eV} \ .
\end{equation}
Now the VBO for CdSe/ZnTe can simply be calculated as
\begin{equation}
\Delta E_\txt{v} = E_\txt{g}(\txt{CdSe}) - E_0.
\label{eq:finalbandoffset}
\end{equation}
With the band-gap of cubic CdSe  taken as $E_\txt{g}(\txt{CdSe})=1.76 \ $eV 
\cite{adachi_handbook_2004} the VBO for CdSe/ZnTe results to
\begin{equation}
\Delta E_\txt{v} = (0.75 \pm 0.01) \, \txt{eV}.
\end{equation}

%-------------------------------------------------------------------------
\subsection{Further Comparison Theory-Experiment-Literature}
%-------------------------------------------------------------------------

Our experimentally determined value  of $\Delta E_\txt{v} = (0.75 \pm
0.01)\,\txt{\eV}$ is
larger than the one found in the  measurement 
performed by Yu et al.,~\cite{yu_measurement_1991} where a VBO of 
$\Delta E_\txt{v} = (0.64$ eV $\pm 0.07)$ \eV  is reported (at room
temperature, determined using XPS
measurements) and also larger than the value by Gleim et
al.\,from  valence- and core-level photoelectron
spectroscopy  of $(0.6 \pm 0.1)$ \eV.~\cite{gleim_energy_2002}
The difference between those results and our value for the VBO
 could be a result 
of the different temperatures at which the measurements were performed and in the case
of Yu et al.\, also the 
small thickness of the CdSe layer used.
Our calculated value for the VBO $\Delta E_\txt{v} = (0.7 \pm 0.2)$ \eV is in 
good agreement with the experimental findings presented in this work 
given the theoretical uncertainties of the method.

The determination of the VBO by measuring the type II PL emission is a
very easy and precise method. 
Ostinelli et al. calculated in their work the VBO between InP and
AlGaAsSb-alloys 
of various compositions using the type II transition
\cite{ostinelli_photoluminescence_2006} and come to a similar degree
of precision
for the VBO of these materials. Since 
the measurement of the PL with very high resolution is 
comparably easy, the main sources of error for the VBO
determination lie 
in the evaluation of the PL shift. Here, mainly the range of excitation 
power which can be utilized with a given setup and the model used for the 
fitting determine the overall precision of the VBO calculation. Therefore,  
a good model of the type II emission is very important, not only to 
improve the precision of VBO calculations but to improve the understanding 
of the carrier dynamics at the type II interface in general. Of course, 
the accuracy of
the literature values for the band gap of the constituents is also important, 
as they enter the analysis, see Eq.\,(\ref{eq:finalbandoffset}).

From the theoretical side, it should be noted that our usage of four $sp^3$
bonding-orbitals per Bravais lattice site of the
zincblende structure represents a somewhat natural choice for the calculation
of
BPs.
A detailed microscopic analysis of the bond structure across the surface as
\eg given in Ref.\,\onlinecite{lambrecht_interface-bond-polarity_1990} shows
that the BP can be
identified with the average $sp^3$ hybrid level of the material. The
dielectric screening of the charge transfer across the interface corresponds
to a screening of the difference between these levels, which corresponds
to an alignment of the branch points. Harrison and Tersoff pointed
out that the more the lattice constants of the tetrahedral
semiconductors are alike, the more accurate is this estimation of the band
alignment.~\cite{harrison_tight-binding_1986} The lattice
constants of the cubic ZnTe/CdSe system
under consideration differ by less than 0.2\%, which may explain why this
conceptionally rather simple theoretical approach gives especially good results
in this case.

%%%%%%%%%%%%%%%%%%%%%%%%%%%%%%%%%%%%%%%%%%%%%%%%%%%%%%%%%%%%%%%%%%%%%%%%%%%%%%%

\section{Conclusion and outlook}

In this paper, we have determined the valence band offset (VBO) of the
CdSe/ZnTe
type II heterojunction with underlying zincblende structure. As literature
values for the VBO often show a huge spread,  we have used a theoretical as
well as an experimental approach to crosscheck our results.

On the experimental side, a model of the
interface carrier dynamics has been used to extract an accurate value for the
VBO from the excitation-power dependent photoluminescence (PL) signal of the 
type II interface. We have
then obtained the value $\Delta E_\txt{v} = (0.75 \pm 0.01)$ \eV for the
CdSe/ZnTe VBO, where the accuracy is essentially limited by the
number of data points available from the experiment and the 
literature value of the bulk band gap for cubic bulk CdSe.

On the theoretical side, we have used a refined empirical tight-binding
parametrization for the calculation of the branch point energy of cubic CdSe 
and ZnTe.
The VBO has been determined to $\Delta E_\txt{v} = (0.7 \pm 0.2)$
\eV from the local charge neutrality condition. Here, the accuracy is
predominantly limited by the included one-particle properties of the
constituent materials, but the value confirms the experimental findings
within its error boundaries. 

As theoretical considerations from the literature suggest only a weak
dependence on the interface polarity and orientation, we can in summary
recommend the experimentally determined value of $\Delta E_\txt{v} = (0.75 \pm
0.01)$ \eV as zero-temperature  VBO for the cubic CdSe/ZnTe heterojunction.

Our combined experimental and theoretical approach can in principle be applied
to other type II heterojunctions. In case of systems with a lower
crystal symmetry (e.g.\,wurtzite structure) and/or significantly larger lattice
mismatch, possible dipole corrections and orientational dependencies may have
to be considered on the theoretical side.
On the experimental side, such conditions could lead to a stronger
confinement of the charge carriers 
at the type II interface, in favor of the formation of excitons. In that case,
the 
model of the carrier dynamics has to be corrected accordingly. In all cases, a 
careful analysis of the charge carrier dynamics at the interface is crucial 
to determine the VBO from power dependent PL measurements.

%%%%%%%%%%%%%%%%%%%%%%%%%%%%%%%%%%%%%%%%%%%%%%%%%%%%%%%%%%%%%%%%%%%%%%%%%%%%%%%

\begin{acknowledgments}
 
Daniel Mourad would like to thank Gerd Czycholl and Paul Gartner for fruitful discussions. 
Jan-Peter Richters would like to acknowledge financial support from
CEA/DSM-energie.

\end{acknowledgments}

%%%%%%%%%%%%%%%%%%%%%%%%%%%%%%%%%%%%%%%%%%%%%%%%%%%%%%%%%%%%%%%%%%%%%%%%%%%%%%%

\appendix

\section{Tight-binding parametrization 
\label{sec:tightbindingparametrization}}

Our ETBM parametrization for the zincblende structure uses the same basis set
as Loehr,\cite{loehr_improved_1994} \ie a localized Wannier-type $sp^3$
basis per spin direction:
\begin{equation}\label{eq:sp3basis}
\left| \mathbf{R} \, \alpha \right\rangle, \quad \alpha \in \left\lbrace
s,p_x,p_y,p_z \right\rbrace \times \left\lbrace
\uparrow, \downarrow \right\rbrace.
\end{equation}
Here, $\mathbf{R}$ labels the sites of the fcc lattice, which is the
underlying Bravais lattice of the zincblende structure. We
basically follow the
approach as described in Ref.\,\onlinecite{loehr_improved_1994}, but with
some modifications:
\begin{enumerate}
  \item We do not use the two-center approximation.
  \item In addition to the inclusion of second-nearest neighbors, we also
        include non-vanishing CB hopping matrix elements to $\f{R} = a/2
        \,(2,2,0)$ and equivalent points, where $a$ is the
        conventional lattice constant. This enables us to fit the CB to
        the $L$-point. The inclusion of the actual third-nearest neighbors
        at $\f{R} = a/2\,(2,1,1)$ and equivalent points would in principle
        also be possible. However, we prefer the first choice: It prevents
        an erroneous CB dispersion along $K$--$\Gamma$ that we do not wish to
        fix by a fit to the $U,K$ point, as the corresponding
        energies are rarely known for  most material systems.

  \item The CB and the VBs are decoupled by setting all hopping matrix
        elements (ME) between $s$ and $p$ orbitals to zero. 
  \item The HH/LH VBs are not fitted to the Luttinger
        parameter $\gamma_3$, \ie we do not fit to the corresponding
        zone center effective masses along the [111] direction.\footnote{
        Points 3 and 4 could render this parametrization less suitable 
        for the calculation of optical properties than Loehr's scheme. For our
        branch point calculations however, the CB-VB coupling and the fit to
        [111] zone center masses can be neglected.}
\end{enumerate}
Using the usual notation
% $E_{\alpha \alpha'}^{\mathbf{R}
%\mathbf{R'}} = \left\langle \mathbf{R} \alpha \right| H \left|
%\mathbf{R'} \alpha' \right\rangle$
by Slater and Koster
\cite{slater_simplified_1954} for the matrix elements, we obtain:
\begin{eqnarray}
E_{ss}^{000} &=& \frac{3}{8} \, \Gamma_{1}^\txt{c} + \frac{1}{4} \,
L_{1}^\txt{c} +
\frac{3}{8} \, X_{1}^\txt{c} + \frac{3}{2}
\frac{\hbar^{2}}{a^{2} m_{c}} , \nonumber\\
E_{ss}^{110} &=& \frac{1}{16} \, \Gamma_{1}^\txt{c}  - \frac{1}{16}\,
X_{1}^\txt{c}, \nonumber\\
E_{ss}^{200} &=& \frac{1}{48} \, \Gamma_{1}^\txt{c} - \frac{1}{12} \,
L_{1}^\txt{c} +
\frac{1}{16} \, X_{1}^\txt{c} \nonumber\\
E_{ss}^{220} &=& -\frac{1}{48}\,\Gamma_{1}^\txt{c} + \frac{1}{48}\,
L_{1}^\txt{c}- \frac{1}{8}
\frac{\hbar^{2}}{ a^{2} m_{c}}, \nonumber\\
E_{xx}^{000} &=& \frac{5}{8} \, \Gamma_{15}^\txt{v} + \frac{1}{8} \,
X_{3}^\txt{v}  + \frac{1}{4} \, X_{5}^\txt{v}  - 3 \frac{\hbar^{2}}{ a^{2}
m_{0}} \gamma_1 + 4 \frac{\hbar^{2}}{ a^{2} m_{0}} \gamma_2 ,\nonumber\\
E_{xx}^{110} &=& \frac{1}{16} \, \Gamma_{15}^\txt{v} - \frac{1}{16} \,
X_{3}^\txt{v}, \label{eq:3plus1ebom}\\
E_{xx}^{011} &=& \frac{1}{16} \, \Gamma_{15}^\txt{v} + \frac{1}{16} \,
X_{3}^\txt{v} -
\frac{1}{8} \, X_{5}^\txt{v}, \nonumber\\
E_{xx}^{200} &=& -\frac{1}{16}\Gamma_{15}^\txt{v} + \frac{1}{16}\,X_{3}^\txt{v}
+
\frac{1}{2} \frac{\hbar^{2}}{ a^{2} m_{0}} \gamma_1, \nonumber\\
E_{xx}^{002} &=& -\frac{1}{16}\Gamma_{15}^\txt{v} + \frac{1}{16}\,
X_{5}^\txt{v} + \frac{1}{2}\, \frac{\hbar^{2}}{ a^{2} m_{0}} \gamma_{1} - 2
\frac{\hbar^{2}}{ a^{2} m_{0}} \gamma_{2}, \nonumber\\
E_{xy}^{110} &=& -\frac{1}{4} \, \Gamma_{15}^\txt{v}  +\frac{1}{4}
L_{3}^\txt{v} +
\frac{3}{2} \frac{\hbar^{2}}{a^{2} m_{0}} \gamma_1 - 2 \frac{\hbar^{2}}{ a^{2}
m_{0}} \gamma_2. \nonumber
\end{eqnarray}

Here, $m_0$ is the free electron
mass. The meaning of the remaining input
parameters and the used values can be found in Table
\ref{tab:materialparameters}. All values are
taken from  Ref.\,\onlinecite{adachi_handbook_2004}. The only exceptions are
the Luttinger parameters for zb-CdSe, which are taken from
Ref.\,\onlinecite{kim_optical_1994}, and the lattice constant of ZnTe to match
the value as described in this paper for the present growth conditions.
\footnote{The corresponding set of
Luttinger parameters from Ref.\,\onlinecite{adachi_handbook_2004} were
estimated from a plot of the band gap versus the $\gamma_i$ for cubic
semiconductors. It might be suitable for effective mass and
$\f{k}\cdot \f{p}$ models but leads to erroneous band curvatures for
large $\f{k}$.}
We chose low-temperature input data where available (\eg for the band gap) to
ensure comparability with the experimental boundary conditions.

%-----------------------------------------------------------
\begin{table}
\centering
\caption{Input material parameters for zb-CdSe and
ZnTe. The double group notation is added
in brackets if the corresponding energy values are identical.}
\label{tab:materialparameters}
\begin{tabular}{l|lr|l|l}
\hline
\hline
Parameter& Description & & CdSe &  ZnTe \\
\hline
$a$ & lattice constant & (\AA) &  \phantom{+}6.077  &
 \phantom{+}6.089 \\
$\Delta_\txt{SO}$ & spin-orbit splitting & (\eV) &  \phantom{+}0.41
 & \phantom{+}0.95\\
$\gamma_1$ & Luttinger parameter &{}&  \phantom{+}3.33 
 & \phantom{+}3.96
\\
$\gamma_2$ & Luttinger parameter &{}& \phantom{+}1.11  &  \phantom{+}0.86\\
$m_\txt{c}$ & CB effective mass & ($m_0$) &  \phantom{+}0.119 
 & \phantom{+}0.11\\

$\Gamma_{1}^\txt{c}\,(\Gamma_{6}^\txt{c})$ & CB energy  &
(\eV) &  \phantom{+}1.76 
&  \phantom{+}2.38\\
$\Gamma_{15}^\txt{v}$ & HH/LH VB energy &
(\eV) & \phantom{+}0 
 &  \phantom{+}0 \\
$X_1^\txt{c}\,(X_6^\txt{c})$ & CB energy & (\eV) & 
\phantom{+}4.37& \phantom{+}3.05\\
$X_5^\txt{v}$ & HH/LH VB energy& (\eV) & $-$1.78 &$-$2.4
\\
$X_3^\txt{v}$ &split-off VB energy  & (\eV) & $-$4.0 &$-$5.2 \\
$L_1^\txt{c} (L_6^\txt{c})$ &  CB energy  & (eV) & \phantom{+}3.87 
&\phantom{+}3.07 \\
$L_3^\txt{v}$  & HH/LH VB energy  & (eV) & $-$0.71 & $-$1.1 \\
\hline
\hline
\end{tabular}
\end{table}
%---------------------------------------------------------------------

Note that the system of
equations (\ref{eq:3plus1ebom}) still uses the usual single group notation for
the corresponding energy values. The addition of a spin-orbit coupling
Hamiltonian $H_\txt{SO}$ 
on the same level of approximation as in Ref.\,\onlinecite{loehr_improved_1994}
lifts
degenerations. The topmost VB with single group symmetry
$\Gamma_{15}^\txt{v}$ is split into a fourfold state $\Gamma_{8}^\txt{v}$ and a
twofold state $\Gamma_{7}^\txt{v}$, where
\begin{equation}
 \Gamma_{8}^\txt{v} - \Gamma_{7}^\txt{v} = \Delta_\txt{SO}
\end{equation}
is the spin-orbit splitting constant.
Similarly, the spin-orbit coupling splits the $X_{5}^\txt{v}$ edge to
\begin{equation}
X_{7}^\txt{v} = X_{5}^\txt{v} + \Delta_\txt{SO} / 3
\end{equation}
for the HH band. These
analytically calculable linear shifts are being included explicitly into the
set of equations
(\ref{eq:3plus1ebom}) by substitution. While the
conduction band energies $\Gamma_{1}^\txt{c} (= \Gamma_{6}^\txt{c})$,
$X_1^\txt{c} (=X_6^\txt{c})$ and $L_1^\txt{c} (=L_6^\txt{c})$ remain unaffected
by the spin, the shifts $X_5^v \rightarrow X_6^v$ (LH), $L_3^\txt{v}
\rightarrow L_{4,5}^\txt{v}$ (HH) and $L_3^\txt{v} \rightarrow L_{6}^\txt{v}$
(LH) stem from higher-order roots of the characteristic polynomial and are
only 
included implicitly via
$H_\txt{SO}$.

As usual, the band structure $E_n(\f{k})$ is then obtained by
diagonalization of the tight-binding matrix $\sum_{\mathbf{R}} \exp{(i
\mathbf{k} \cdot \mathbf{R}) } E_{\alpha  \alpha'}^{lmn}$
%
% \begin{equation}
%     \sum_{\mathbf{R}} e^{i \mathbf{k} \cdot \mathbf{R} }
%     E_{\alpha  \alpha'}^{\mathbf{0} \mathbf{R}}
% \label{eq:tbmatrix}
% \end{equation}
%
for each $\f{k}$ in the irreducible BZ.

%%%%%%%%%%%%%%%%%%%%%%%%%%%%%%%%%%%%%%%%%%%%%%%%%%%%%%%%%%%%%%%%%%%%%%%%%%%%%

\section{Estimation of the interface dipole contribution 
\label{sec:dipolecontribution}}

To estimate the contribution from interface dipoles, we closely follow
Refs.\,\onlinecite{monch_empirical_1996,monch_electronic_2004}. There, the VBO
including dipole corrections is given as
\begin{equation}
\Delta E_\txt{v} = E_\txt{BP}^\txt{B} - E_\txt{BP}^\txt{A} + D_x(X_B -
X_A).
\end{equation}
\label{eq:dipolecorrection}
Here $X_A$ and $X_B$ are electronegativity values from the Miedema scale and 
\begin{equation}
D_x = \frac{A_x}{1+0.1(\varepsilon_\infty-1)^2},
\end{equation}
\label{eq:Miedemacorrection}
where $A_x = 0.86$ \eV/Miedema-unit is the proportionality factor between the
work function and the electronegativity, while $\varepsilon_\infty$ is the
optical dielectric constant of the semiconductor.
The $X_{A/B}$  for
binary compound semiconductors are obtained as the geometric mean of their
constituents' values.

Using the data from Table A.4 of
Ref.\,\onlinecite{monch_semiconductor_2001}, we obtain $X_\txt{CdSe} = (4.05
\times 5.79)^{1/2} \approx 4.84$ and $X_\txt{ZnTe} =(4.10 \times 4.92)^{1/2}
\approx 4.49$.  Here, the Miedema electronegativity values for Se and Te 
have been obtained from the corresponding Pauling values by the approximate conversion 
$X_\txt{Mied} \approx 1.93 \, X_\txt{Paul} + 0.87$,
 see e.\,g.\,Eq.\,(5.19) of Ref.\,\onlinecite{monch_electronic_2004}.
The low-temperature dielectric constants for zb-CdSe and ZnTe
are approximately 7.8 and 6.7.~\footnote{Please note that the value of
$\varepsilon_\infty =6.2$ from Adachi's data collection,
Ref.\,\onlinecite{adachi_handbook_2004}, 
is not suitable here, as it is a directionally averaged result from wurtzite phase data
at room temperature. In particular, the dielectric response should become larger when the bulk band gap becomes smaller. A possible approach to find consistent values for II-VI semiconductors is the use 
of the empirical Moss model as done by Gupta and Ravindra in
Ref.\,\onlinecite{gupta_comments_1980}. 
This has also succesfully been applied to further II-VI systems, see
Ref.\,\onlinecite{mourad_multiband_2010-1}. Our value for CdSe is also
in satisfactory agreement with recent values in the Landolt-B\"{o}rnstein
database.} We use the 
first value, as the ZnTe valence electrons exponentially decay into the
CdSe barrier. Then, the correction term
$D_x(X_\txt{CdSe} - X_\txt{ZnTe})$ is approximately \mbox{+0.05 \eV}. This
value just falls in the accuracy range of the charge neutrality condition as
mentioned in Sec.\,\ref{subsec:calculationofvbo}.

Although the calculated and the measured VBO for the CdSe/ZnTe junction would coincide
almost perfectly when this dipole term is taken into account, we
refrain from doing so (as common in the literature for junctions between
isovalent semiconductors)\cite{monch_electronic_2004}. First, additional
empirical parameters and assumptions enter the calculation of this
contribution. Second, it
would pretend a false degree of accuracy, as the calculation of the branch
point itself is subject to several sources of uncertainties, as mentioned
throughout this paper.

%%%%%%%%%%%%%%%%%%%%%%%%%%%%%%%%%%%%%%%%%%%%%%%%%%%%%%%%%%%%%%%%%%%%%%%%%%%%%%

%\vspace{2cm}
% WARNING: No whitespace after comma when including multiple bib-files ;-)
%\bibliography{references_jan,references_daniel}{}
%\bibliographystyle{apsrev}

\end{document}